\documentclass[11pt]{article}
\usepackage{sigsam, amsmath}

\usepackage{hyperref}

\issue{}
\articlehead{}
\usepackage{xcolor}

\begin{document}

\titlehead{The DEWCAD Project}
\authorhead{R.Bradford et al.}

\title{The DEWCAD Project:  Pushing Back the Doubly Exponential Wall of Cylindrical Algebraic Decomposition}

\author{
R.~Bradford$^{1}$,
J.H.~Davenport$^{1}$,  
M.~England$^{2}$,  
A.~Sadeghimanesh$^{2}$, 
A.Uncu$^{1}$
\\ \qquad \\
$^1$University of Bath, U.K. \\
\texttt{\{R.Bradford, J.H.Davenport, aku21\}@bath.ac.uk}\\
$^2$Coventry University, U.K.\\
\texttt{\{Matthew.England, Amirhossein.Sadeghimanesh\}@coventry.ac.uk}
}

\date{}

\maketitle

\begin{abstract}
This abstract seeks to introduce the ISSAC community to the DEWCAD project, which is based at Coventry University and the University of Bath, in the United Kingdom.  The project seeks to push back the Doubly Exponential Wall of Cylindrical Algebraic Decomposition, through the integration of SAT/SMT technology, the extension of Lazard projection theory, and the development of new algorithms based on CAD technology but without producing CADs themselves.  The project also seeks to develop applications of CAD and will focus on applications in the domains of economics and bio-network analysis.
\end{abstract}

\section{Cylindrical Algebraic Decomposition}

\subsection{Introduction}
 
\emph{Cylindrical Algebraic Decomposition} (CAD), was developed by Collins in the 1970s.   A CAD is a \emph{decomposition} of ordered $\mathbb{R}^n$ space into cells (i.e. connected subsets) which are \emph{semi-algebraic} so each is described by a sequence of polynomial constraints.  Each constraint involves one further variable (locally cylindrical cells) meaning that the bounds of cells can be read easily.  For any pair of cells the projections with respect to the given ordering are either equal or disjoint (global cylindricity).  See \cite{CJ98} for a detailed exposition.

A CAD is produced relative to Tarski formulae, i.e. logical formulae whose atoms are sign conditions on non-linear polynomial constraints with integer coefficients.  The CAD is produced to be \emph{truth-invariant} for the formulae, commonly achieved through being sign-invariant for the polynomials involved.  A CAD may then be used to give intuitive descriptions of the solutions of such formulae, and to solve associated problems like real Quantifier Elimination (given a quantified formulae produce an unquantified one which is logically equivalent).  

CAD has a large range of potential applications, such as proving collisions of autonomous vehicles impossible \cite{ST11b}; artificial intelligence to pass a university entrance exam \cite{AMIA14}; the derivation of optimal numerical schemes \cite{EH16}; and structural design to minimise the weight of trusses \cite{CC18}.

However, CAD has worst case complexity doubly exponential in the number of indeterminates (both quantified and unquantified variables) \cite{BD07}, meaning that as problem sizes rise you inevitably \textit{hit the doubly exponential wall}.  It is thanks to over 40 years of extensive research that it may be used for applications like those above.  
We seek to push the doubly exponential wall back further still to bring new applications within scope.

\newpage

\subsection{Recent Developments}

For most of its history CAD has followed an algorithmic framework of projection (identifying key polynomials needed to form the decomposition) and lifting (constructing the cells incrementally by dimension). But in recent years there have been a variety of alternative frameworks proposed: e.g. cylindrical decompositions of complex space which are then refined to CADs \cite{CMXY09}; exploiting Boolean structure in the algebraic procedures \cite{BDEMW16}, \cite{EBD20}; \cite{Strzebonski2016}.  

This has led to the development of algorithms and structures which relax the definition of CAD itself.  Non-uniform CAD (NuCAD) maintains the locally cylindrical cell descriptions but relaxes the global cylindricity condition of cells being arranged in cylinders \cite{Brown2015}.  Cylindrical Algebraic Coverings (CACs) \cite{ADEK21} produce cells which are arranged cylindrically but may overlap (so they form a covering, not a decomposition).  Both offer significant savings over an actual CAD.

These new approaches have been inspired by the search based algorithms employed by SAT/SMT solvers, most notably the NLSAT algorithm \cite{JdM12} which determines satisfiability of Tarski formulae through a combination of Boolean search and the construction of cylindrical cells to rule out portions of the search space generalised from failure at a model point.  In turn, the SMT community \cite{BSST09} has started to show interest in the algorithms of computer algebra which may be adapted for their solvers.  The SC$^2$ initiative seeks to draw these communities together \cite{AAB+16a}.

\section{The DEWCAD Project}

The DEWCAD project is funded by the UK's Engineering and Physical Sciences Research Council.  It runs 2021$-$2025 and employs the authors, at Coventry University and the University of Bath in the United Kingdom.

\subsection{Our Research Objectives}

Our initial aim is the implementation of CAD infrastructure in Maple that can support both the use of CAD for traditional quantifier elimination, and also CAD as a theory solver for SMT as in \cite{KA20}.  We hypothesise that it will be more efficient to integrate SAT into computer algebra that the reverse, given (a) the complexity of algebraic procedures and the potential to benefit from decades of development of underlying sub algorithms and (b) the software engineering traditions within the SAT community that allow for standardised I/O and easier code reuse.

Our next objective is to reuse that infrastructure to implement the CAD-like algorithms discussed above \cite{JdM12}, \cite{Brown2015}, \cite{ADEK21} in Maple.  These approach all relax the requirements of CAD in different ways and it is not clear which is superior:  comparing them in a common system will allow for more meaningful conclusions on the underlying algorithms.  There is also substantial scope for theory development on all such algorithms.  Further, the experiments in \cite{ADEK21} suggested there are substantial sets of problems on which each of the algorithms may excel giving rise to the potential of a portfolio solver (analogous to say \cite{XHHL08}) and the use of machine learning for this and other choices \cite{England2018}.

CAD projection identifies those polynomials in less variables whose zeros represent changes in behaviour of the input polynomials.  Simplifications to CAD projection have been critical to historic improvements.  For a long time the best (i.e. smallest) operator was that of McCallum \cite{McCallum1998}, which has been developed into a family of operators specialised to logical structure in the input e.g. \cite{McCallum1999b}, \cite{BDEMW16}, \cite{EBD20}.  However, all members of this family could fail for a small class of input types.  Recently, the Lazard projection\footnote{first suggested in the 1990s by Lazard but verified recently by McCallum and collaborators \cite{MH16}, \cite{MPP19}, \cite{BM20}} has been shown to avoid such failure while being no more expensive than \cite{McCallum1998} except on cases where it failed.  However, it remains to extend this theory into the wider family, with the initial work of \cite{NDS19}, \cite{NDS20} to be continued in this project.

\subsection{New Application Domains}

As noted above, CAD has a great many applications throughout the sciences and engineering. The DEWCAD project seeks to focus in detail on two emerging ones.

Bio-chemical network analysis has seen increasing use of computer algebra techniques, with recent examples including \cite{BDEEGGHKRSW20} and \cite{RS21}. In both of these studies it was a combination of CAD with other algebraic techniques that allowed for an exact solution. Also, in both studies a comparison with solutions from numerical methods was made which showed the potential for numerical methods to make errors through floating point accumulation or insufficient sampling.

Recently CAD has also been shown to have use within economics \cite{MBDET18} \cite{MDE18}. These range from the educational (Chicago now uses QE
technology in its economics undergraduate curriculum) to the topical (QE used to demonstrate a gap in the reasoning of a Nobel Laureate).

\subsection{Project partners}

The DEWCAD project plans to work with a range of partners:  the software company Maplesoft on implementation within Maple; \'{A}br\'{a}ham (RWTH Aachen) on integration with SMT; Brown (US Naval Academy) on search based algorithms;  McCallum (Macquarie University) on the development of Lazard projection; Mulligan (U. Chicago) on economics applications; the SYMBIONT project \cite{BFRSSSSWW19} on bio-chemical network applications.  We welcome collaboration with other members of the ISSAC community who have an interest in any of the topics described here or other potential CAD applications that may come into scope if the doubly exponential wall were pushed back further still.

\section*{Acknowledgements}

Bradford, Davenport and Uncu acknowledge the support of EPSRC Grant EP/T015713/1, while England and Sadeghimanesh acknowledge the support of EPSRC Grant EP/T015748/1.


\end{document}